%% file: lensing-lfi.tex
\documentclass[twocolumn]{aastex63}

\usepackage{amsmath, amsthm, amssymb, amsfonts, xspace}
\usepackage{fontawesome}

\input{definitions}

\shorttitle{Inferring dark matter substructure with machine learning}
\shortauthors{Brehmer and Mishra-Sharma et al.}

\begin{document}\sloppy\sloppypar\raggedbottom\frenchspacing

\title{\textbf{
Mining for Dark Matter Substructure: \\
Inferring subhalo population properties from strong lenses with machine learning
}}

\correspondingauthors{Siddharth Mishra-Sharma}{Johann \newline Brehmer}
\email{sm8383@nyu.edu}{johann.brehmer@nyu.edu}

\author[0000-0003-3344-4209]{Johann Brehmer}
\altaffiliation{Both authors contributed equally to this work}
\affiliation{Center for Cosmology and Particle Physics, Department of Physics, New York University, 726~Broadway, New York, NY 10003, USA}
\affiliation{Center for Data Science, New York University, 60 Fifth Ave, New York, NY 10011, USA}

\author[0000-0001-9088-7845]{Siddharth Mishra-Sharma}
\altaffiliation{Both authors contributed equally to this work}
\affiliation{Center for Cosmology and Particle Physics, Department of Physics, New York University, 726~Broadway, New York, NY 10003, USA}

\author[0000-0002-1471-2063]{Joeri Hermans}
\affiliation{Montefiore Institute, University of Li\`ege, Belgium}

\author[0000-0002-2082-3106]{Gilles Louppe}
\affiliation{Montefiore Institute, University of Li\`ege, Belgium}

\author[0000-0002-5769-7094]{Kyle Cranmer}
\affiliation{Center for Cosmology and Particle Physics, Department of Physics, New York University, 726~Broadway, New York, NY 10003, USA}
\affiliation{Center for Data Science, New York University, 60 Fifth Ave, New York, NY 10011, USA}

\begin{abstract}\noindent
The subtle and unique imprint of dark matter substructure on extended arcs in strong lensing systems contains a wealth of information about the properties and distribution of dark matter on small scales and, consequently, about the underlying particle physics. However, teasing out this effect poses a significant challenge since the likelihood function for realistic simulations of population-level parameters is intractable. We apply recently-developed simulation-based inference techniques to the problem of substructure inference in galaxy-galaxy strong lenses. By leveraging additional information extracted from the simulator, neural networks are efficiently trained to estimate likelihood ratios associated with population-level parameters characterizing substructure. Through proof-of-principle application to simulated data, we show that these methods can provide an efficient and principled way to simultaneously analyze an ensemble of strong lenses, and can be used to mine the large sample of lensing images deliverable by near-future surveys for signatures of dark matter substructure. \href{https://github.com/smsharma/StrongLensing-Inference}{\faGithub}
\end{abstract}

\keywords{
astrostatistics techniques (1886)
---
cosmology (343)
---
dark matter (353)
---
gravitational lensing (670)
---
nonparametric inference (1903)
---
strong gravitational lensing (1643)
}

\tableofcontents{}

\section{Introduction}
\label{sec:intro}

Dark matter (DM) accounts for nearly a quarter of the energy budget of the Universe, and pinning down its fundamental nature and interactions is one of the most pressing problems in cosmology and particle physics today. Despite an organized effort to do so through terrestrial~\citep[\eg,][]{2017PhRvL.118b1303A,2017PhRvL.119r1302C,2018PhRvL.121k1302A}, astrophysical~\citep[\eg,][]{2017ApJ...834..110A,2018PhRvD..98l3004C,2018PhRvL.120j1101L}, and collider~\citep[\eg,][]{2017PhLB..769..520S,2019arXiv190301400A} searches, no conclusive evidence of interactions between the Standard Model (SM) and dark matter exists to-date.

Meanwhile, dark matter can also be studied directly through its irreducible gravitational interactions. The concordance $\Lambda$ Cold Dark Matter (\lcdm) framework of non-relativistic, collisionless dark matter particles provides an excellent description of the observed distribution of matter on large scales. However, many well-motivated models predict deviations from \lcdm on smaller scales. Fundamental dark matter microphysical properties such as its particle mass and self-interaction cross-section can imprint themselves onto its macroscopic distribution in ways that can be probed by current and future experiments~\citep{1712.06615,2019arXiv190201055D,1903.04742}. As a motivating example, early decoupling of relativistic dark matter species from the cosmic plasma would cause it to have a significant free-streaming length, leading to an underabundance of lower-mass subhalos today~\citep{1983ApJ...274..443B,2001ApJ...556...93B,astro-ph/0004381,0807.0622,1008.0992}. Dark matter self-interactions~\citep{astro-ph/9909386,astro-ph/0006134,astro-ph/0006218,astro-ph/0205322,1201.5892,1208.3026,1211.6426,1311.6524,1508.03339,1611.02716,1609.08626,1805.03203,1904.10539,1901.00499,1903.01469} and dissipative dynamics in the dark sector~\citep{1303.1521,1702.05482,1706.04195,1707.03829} are examples of scenarios that would modify the structure of the subhalo density profiles in addition to possibly depressing the abundance of lower-mass halos as compared to CDM predictions in the latter case~\citep{1405.2075,1412.4905,1512.05349}.

There exist several avenues for probing the distribution of dark matter on small scales. While the detection of ultrafaint dwarf galaxies through the study of stellar overdensities and kinematics~\citep{0706.2687,1503.02079,1503.02584,1508.03622} can be used to make statements about the underlying dark matter properties, theoretical uncertainties in the connection between stellar and halo properties~\citep{1809.05542,1804.03097} and the effect of baryons on the satellite galaxy population~\citep{1608.01849,1701.03792,1812.00044,1811.11791} pose a challenge. Furthermore, suppressed star formation in smaller halos means that there exists a threshold ($\lesssim 10^8\,\mathrm{M}_\odot$) below which subhalos are expected to be mostly dark and devoid of baryonic activity~\citep{1992MNRAS.256P..43E,1611.02281,1607.03127}. This makes studying the imprint of gravitational interactions the \emph{only} viable avenue for probing substructure at smaller scales. In this spirit, the study of subhalo-induced perturbations to the kinematic phase-space distribution in cold stellar streams~\citep{astro-ph/9807243,1109.6022,1303.4342,1804.06854,1811.03631}, and in Galactic stellar fields~\citep{1711.03554} have been proposed as methods to look for low-mass subhalos through their gravitational interactions in the Milky Way.

Complementary to the study of locally-induced gravitational effects, gravitational lensing has emerged as an important technique for studying the distribution of matter over a large range of scales. Locally, the use of time-domain astrometry has been proposed as a promising method to measure the distribution of local substructure through correlated, lens-induced motions of background celestial objects due to foreground subhalos~\citep{2018JCAP...07..041V}. In the extragalactic regime, galaxy-scale strong lenses are laboratories for studying dark matter substructure. The typical substructure abundance within galaxy-scale lenses has been constrained through the measurement of positions and flux ratios of multiple images in quasar lenses~\citep{2002ApJ...572...25D,2019arXiv190504182H} and lensed images of extended~\citep{0910.0760,1002.4708,1201.3643,1601.01388} as well as quasar sources~\citep{1109.0548,1402.1496,1701.05188,1908.06983} have been used to set limits on the abundance of or find evidence for individual subhalo clumps with masses $\gtrsim 10^8\,\mathrm{M}_\odot$. Although these individual high-significance detections can be used to derive constraints on substructure abundance and the subhalo mass function, searches for one (or a general fixed number of) subhalos do not take into account covariances between models with different numbers of subhalos and can leave unexpressed the degeneracies between, e.g., the imprint of several low-mass subhalos and that of a massive subhalo perturber. Additionally, these detections by definition probe the most massive subhalos in the lensing galaxies which, given the particle physics-motivated goal of constraining small-scale structure, is the less interesting regime compared to probing the fainter end of the subhalo mass function.

Another approach relies on probing the collective effect of sub-threshold (\ie, not individually resolvable) subhalos on extended arcs in strongly lensed systems. A particular challenge here is that the  properties of the individual subhalos correspond to a high-dimensional space of latent variables, which must be marginalized to compute the likelihood. This complicated marginalization integral makes the likelihood for population-level parameters effectively intractable. Methods based on summary statistics~\citep{1702.00009} and studying the amplitude of spatial fluctuations on different scales through a power spectrum decomposition~\citep{1403.2720,1506.01724,1707.04590,1710.03075,1809.00004,1806.07897,1808.03501} have been proposed as ways to reduce the dimensionality of the problem and enable substructure inference in a tractable way. This class of methods is well-suited to studying dark matter substructure since they can be sensitive to the \emph{population} properties of low-mass subhalos in strongly lensed galaxies which are directly correlated with the underlying dark matter particle physics.

Particularly promising in this regard are trans-dimensional techniques like probabilistic cataloging~\citep{1508.00662,1706.06111} that have been proposed to take into account covariances between models with different numbers of subhalos in a principled manner and can efficiently map out the parameter space associated with multiple sub-threshold objects in lensing systems. The output of such analyses is a ensemble of posterior-weighted subhalo catalogs which can be marginalized over to infer higher-level parameters (hyperparameters) characterizing the population properties of subhalos, potentially over multiple lensing images. These results can be highly sensitive to the assumed metamodel complexity however~\citep{1706.06111} and potentially computationally limited for a large number of lenses as they require running an independent analysis to produce a probabilistic catalog for each image.

Current and near-future observatories like DES~\citep{1601.00329}, LSST~\citep{0912.0201,2019arXiv190201055D,1902.05141}, and \Euclid~\citep{1001.0061} are expected to find hundreds to thousands of galaxy-galaxy strong lenses~\citep{1001.2037,1003.5567,2015ApJ...811...20C}, making substructure inference in these systems (and high-resolution followups on a subset) one of the key avenues for investigating dark matter substructure and stress-testing the Cold Dark Matter paradigm in the near future. This calls for methods that can efficiently analyze large samples of lensed images to infer the underlying substructure properties with minimal loss of information stemming from dimensional reduction.

In recent years, a large number of methods have been developed that train neural networks to estimate the likelihood function, likelihood ratio function, or posterior~\citep{2012arXiv1212.1479F, 2014arXiv1410.8516D, 2015arXiv150203509G, 2015arXiv150505770J, Cranmer:2015bka,  2016arXiv160508803D, 2016arXiv160206701P, NIPS2016_6084, 2016arXiv161110242D, 2016arXiv160502226U,  2016arXiv160605328V, 2016arXiv160903499V, 2016arXiv160106759V, 2017arXiv170208896T, 2017arXiv170507057P, 2017arXiv170707113L, 2017arXiv171101861L, gutmann2017likelihood, DBLP:journals/corr/abs-1806-07366, 2018arXiv181009899D, 2018arXiv181001367G, 2018arXiv180400779H, 2018arXiv180703039K, 2018arXiv180509294L, 2018arXiv180507226P, Alsing:2019xrx, Hermans:2019ioj}. These techniques can be directly applied to population-level parameters, avoiding an additional marginalization step. In contrast to traditional simulation-based (or ``likelihood-free'') approaches, namely Approximate Bayesian Computation, they do not rely on summary statistics and instead learn to extract information directly from the full input data, which in our case corresponds to the observed lensed images. Finally, some of these methods let us to amortize the computational cost of the inference---after an upfront simulation and training phase, inference for any observed lensed image is efficient, enabling a simultaneous analysis of a large number of observations.

In this paper, we follow this approach and apply a particularly powerful technique for simulation-based inference introduced in~\citet{1805.00013, 1805.00020, 1805.12244, Stoye:2018ovl} to the problem of extracting high-level substructure properties from an ensemble of galaxy-galaxy strong lensing images. This method extracts additional information from the simulator, which is then used to train a neural network as a surrogate for the likelihood ratio function. The additional information increases the sample efficiency during training and thus reduces the computational cost. A calibration procedure ensures correct inference results even in the case of imperfectly trained networks. We demonstrate the feasibility of this method on a catalog of simulated lenses. After discussing the information content in individual lensed images, we switch to a simultaneous analysis of multiple observed images and calculate the expected combined constraints on population-level substructure parameters in both a frequentist and a Bayesian setup.

This paper is organized as follows. In Section~\ref{sec:lensing-formalism} we briefly review the formalism of gravitational strong lensing and describe our simulation setup, including the assumptions we make about the population of background sources and host galaxies, the substructure population, and observational parameters. In Section~\ref{sec:lfi-formalism} we describe the simulation-based analysis technique used and its particular application to the problem of mining dark matter substructure properties from an ensemble of extended lensed arcs. We show a proof-of-principle application to simulated data in Section~\ref{sec:results} and comment on how this method can be extended to more realistic scenarios in Section~\ref{sec:extensions}. We conclude in Section~\ref{sec:conclusions}. In the spirit of reproducibility, code associated with this paper is available on GitHub \githubmaster and we provide links below each figure (\nbicon) pointing to the \package{Jupyter} notebooks used to generate them.

\section{Strong lensing formalism and simulation setup}
\label{sec:lensing-formalism}

In strong lensing systems a background light source is gravitationally lensed by an intervening mass distribution, resulting in multiple localized images on the lens plane (in the case of a point-like quasar source) or an arc-like image (in the case of an extended galaxy source). The latter provides the ability to probe substructure over a relatively larger region on the lens plane. Additionally, young, blue galaxies are ubiquitous in the redshift regime $z\gtrsim1$ and dominate the faint end of the galaxy luminosity function, resulting in a larger deliverable sample of galaxy-galaxy strong lenses compared to that of multiply-imaged quasars. For these reasons, we focus our method towards galaxy-galaxy lenses---systems with extended background sources producing images with lensed arcs---although the techniques presented here can also be applied to samples of lensed quasars.

We now briefly review the basic mathematical formalism behind strong gravitational lensing before describing in turn the models for the lensing galaxy, background source, and dark matter substructure assumed in this study. We also describe the mock observational parameters assumed for the image sample as well as the population properties of the host lenses. Taken together, these define our lensing forward model. Note that we use natural units with $c=1$ throughout this paper.

\subsection{Strong lensing formalism}

Given a mass distribution with dimensionless projected surface mass density $\kappa(\boldsymbol{\theta})=\Sigma(\boldsymbol{\theta}) / \Sigma_{\mathrm{cr}}$, where 
\begin{equation}
\Sigma_{\mathrm{cr}}\equiv \frac{1}{4 \pi G_\mathrm{N}} \frac{D_{\mathrm{s}}}{D_{\mathrm{ls}} D_{\mathrm{l}}}
\end{equation}
%$\Sigma_{\mathrm{cr}}\equiv (1 / 4 \pi G_\mathrm{N}) \, D_{s} / D_{{ls}} D_{{l}}$ 
is the critical lensing surface density and $D_\mathrm{l}$, $D_{\mathrm s}$, and $D_{\mathrm{ls}}$ are the observer-lens, observer-source, and lens-source angular diameter distances respectively, the two-dimensional projected lensing potential is given by~\citep[\eg,][]{1992grle.book.....S,astro-ph/9912508}
\begin{equation}
\psi(\boldsymbol{\theta})=\frac{1}{\pi} \int \diff \boldsymbol{\theta^\prime}\,\ln |\boldsymbol{\theta}-\boldsymbol{\theta^\prime}|\,\kappa(\boldsymbol{\theta^\prime}) .
\end{equation}
The reduced deflection angle is given by the gradient of the projected lensing potential,
\begin{equation}
\boldsymbol{\phi}(\boldsymbol{\theta}) = \nabla \psi(\boldsymbol{\theta}) = \frac{1}{\pi} \int \diff \boldsymbol{\theta^\prime}\,\frac{\boldsymbol{\theta}-\boldsymbol{\theta^\prime}}{|\boldsymbol{\theta}-\boldsymbol{\theta^\prime}|^2}\,\kappa(\boldsymbol{\theta^\prime})
\label{eq:deflection}
\end{equation}
and can be used to determine the position of the lensed source $\boldsymbol{\theta}$ through the lens equation,
\begin{equation}
\boldsymbol{\beta}=\boldsymbol{\theta}-\boldsymbol{\phi}(\boldsymbol{\theta})
\end{equation}
where $\boldsymbol{\beta}$ is the position of the source. For an extended source profile $f_{\mathrm s}$, the final lensed image $f^\prime_{\mathrm s}$ can be obtained as the source light profile evaluated on the image plane~\citep[\eg,][]{1706.06111},
\begin{equation}
f^\prime_{\mathrm s}(\boldsymbol{\theta}) = f_{\mathrm s}\left(\boldsymbol{\theta}-\boldsymbol{\phi}(\boldsymbol{\theta})\right).
\label{eq:lensed_image}
\end{equation}

Given a lens density profile, the deflection vector can be computed using Equation~\eqref{eq:deflection}, and analytic expressions for many commonly considered profiles are available in the literature~\citep[\eg,][]{2001astro.ph..2341K}. The projected lensing potential and mass density are related through the Poisson equation $\nabla^2 \psi(\boldsymbol{\theta}) = 2\kappa(\boldsymbol{\theta})$, and its linearity implies that the combined projected potential due to multiple perturbers can be written as the sum of individual potentials, and the individual deflections can be superimposed as a consequence. The total deflection can then be used to calculate the lensed image for a given source profile using Equation~\eqref{eq:lensed_image}. For more details on the gravitational lensing formalism see, \eg,~\citet{1992grle.book.....S,astro-ph/9912508,1003.5567}.

\subsection{Lensing host galaxy}

Cosmological $N$-body simulations suggest that the dark matter distribution in structures at galactic scales can be well-described by a universal, spherically symmetric Navarro-Frenk-White (NFW) profile. However, strong lensing probes a region of the host galaxy much smaller than the typical virial radii of galaxy-scale dark matter halo, and the mass budget here is dominated by the baryonic bulge component of the galaxy. Taking this into account, the total mass budget of the lensing host galaxy, being early-type, can be well-described by a singular isothermal ellipsoid (SIE) profile. Since neither the dark matter nor the baryonic components are individually isothermal, this is sometimes known as the bulge-halo conspiracy~\citep{1003.5567}. We consider the spherical simplification of the SIE profile, the singular isothermal sphere (SIS), with the density distribution given by~\citep{1994A&A...284..285K,1003.5567}
\begin{equation}
\rho_\mathrm{SIS}(r)=\frac{\sigma_{v}^{2}}{2 \pi G_\mathrm{N} r^2}
\label{eq:hostprofile}
\end{equation}
where $\sigma_{v}$ is the central 1-D velocity dispersion of the lens galaxy and $q$ is the ellipsoid axis ratio, with $q=1$ corresponding to the SIS profile. The Einstein radius for this profile, defining the characteristic lensing scale, is given by~\citep{1003.5567}
\begin{equation}
\theta_{\mathrm{E}}=4 \pi\sigma_{v}^{2} \frac{D_\mathrm{ls}\left(z_\mathrm{l}, z_\mathrm{s}\right)}{D_\mathrm{s}\left(z_\mathrm{s}\right)} \,,
\label{eq:siethetae}
\end{equation}
where $z_\mathrm{l}$ and $z_\mathrm{s}$ are respectively the lens and source redshifts. We use the cosmology from~\citet{1502.01589} to compute cosmological distances throughout this paper.

The deflection field for the SIE profile is given by~\citep{2001astro.ph..2341K}
\begin{align}
\phi_{x} &=\frac{\theta_\mathrm{E} q}{\sqrt{1-q^{2}}} \tan ^{-1}\left[\frac{\sqrt{1-q^{2}} \theta_x}{\chi}\right] \\
\phi_{y} &=\frac{\theta_\mathrm{E} q}{\sqrt{1-q^{2}}} \tanh ^{-1}\left[\frac{\sqrt{1-q^{2}} \theta_y}{\chi+q^{2} }\right]
\end{align}
with $\chi \equiv \sqrt{\theta_x^2 q^2 + \theta_y^2}$ and we explicitly denote our angular coordinates as $\left\{\theta_x, \theta_y\right\}$.

Although the total galaxy mass (baryons + dark matter) describe the macro lensing field, for the purposes of describing substructure we require being able to map the measured properties of an SIE lens onto the properties of the host dark matter halo. To do this, we relate the central stellar velocity dispersion $\sigma_v$ to the mass $M_{200}$ of the host dark matter halo. \citet{2018ApJ...859...96Z} derived a tight correlation between $\sigma_v$ and $M_{200}$, modeled as
\begin{equation}
\log\left(\frac{M_{200}}{10^{12}\,\Msun}\right) = \alpha + \beta\left(\frac{\sigma_v}{100\,\kmps}\right)
\label{eq:sigma_v_M_200_relation}
\end{equation}
with $\alpha = 0.09$ and $\beta = 3.48$.
We model the host dark matter halo with an NFW profile~\citep{1996ApJ...462..563N,1997ApJ...490..493N}

\begin{equation}
\rho_\mathrm{NFW}(r)=\frac{\rho_\mathrm{s}}{\left(r / r_\mathrm{s}\right)\left(1+r / r_\mathrm{s}\right)^{2}}
\label{eq:rhoNFW}
\end{equation}
where $\rho_\mathrm{s}$ and $r_\mathrm{s}$ are the scale density and scale radius, respectively. The halo virial mass $M_{200}$ describes the total mass contained with the virial radius $r_{200}$, defined as the radius within which the mean density is 200 times the critical density of the universe and related to the scale radius through the concentration parameter $c_{200} \equiv r_{200}/r_\mathrm{s}$. Thus, an NFW halo is completely described by the parameters $\{M_{200}, c_{200}\}$. We use the concentration model from~\citet{2014MNRAS.442.2271S} to derive the halo concentration for a given NFW virial mass.

The spherically-symmetric deflection for an NFW perturber is given by~\citep{2001astro.ph..2341K}
\begin{equation}
\phi_{r}=4 \kappa_\mathrm{s} r_\mathrm{s} \frac{\ln (x / 2)+\mathcal{F}(x)}{x} \,,
\label{eq:nfw_deflection}
\end{equation}
where $x \equiv r/r_\mathrm{s}, \kappa_\mathrm{s}\equiv \rho_\mathrm{s}\,r_\mathrm{s}/\Sigma_\mathrm{cr}$, and
\begin{equation}
\mathcal{F}(x)=\left\{\begin{array}{ll}{\frac{1}{\sqrt{x^{2}-1}} \tan ^{-1} \sqrt{x^{2}-1}} & {(x>1)} \\ {\frac{1}{\sqrt{1-x^{2}}} \tanh ^{-1} \sqrt{1-x^{2}}} & {(x<1)} \\ {1} & {(x=1).}\end{array}\right.
\label{eq:Fx}
\end{equation}

We described the population parameters used to model the host velocity dispersion (and thus its Einstein radius and dark matter halo mass) in Section~\ref{sec:populations} below.

\subsection{Background source}
\label{sec:source}

We model the emission from background source galaxies using a S\'{e}rsic profile, with the surface brightness given by~\citep{1963BAAA....6...41S}
\begin{equation}
f_\mathrm{s}(\theta_r)=f_\mathrm{e} \exp \left\{-b_{n}\left[\left(\frac{\theta_r}{\theta_{r,\mathrm{e}}}\right)^{1 / n}-1\right]\right\},
\end{equation}
where $\theta_{r,\mathrm{e}}$ is the effective circular half-light radius, $n$ is the S\'{e}rsic index, and $b_n$ is a factor depending on $n$ that ensures that $\theta_{r,\mathrm{e}}$ contains half the total intensity from the source galaxy, given by~\citep{1999A&A...352..447C}
\begin{align}
b_n \approx 2 n &- \frac{1}{3} + \frac{4}{405 n} + \frac{46}{25515 n^2} \nonumber \\ &+ \frac{131}{1148175 n^3} - \frac{2194697}{30690717750 n^4}. \nonumber
\end{align}

We assume $n=1$ for the source galaxies, corresponding to a flattened exponential profile and consistent with expectation for blue-type galaxies at the relevant redshifts. $f_\mathrm{e}$ encodes the flux at half-light radius, which can be inferred from the total flux (or magnitude) associated with a given galaxy as follows. For a detector with zero-point magnitude $M_0$, which specifies the magnitude of a source giving 1 count\,s$^{-1}$ in expectation, by definition the total counts are given by $S_\mathrm{tot}=10^{0.4(M-M_0)}$. Requiring the half-light radius to contain half the expected counts, for $n=1$ we have the relation $f_\mathrm{e} \approx 0.526\,t_\mathrm{exp}S_\mathrm{tot} /(2\pi \theta_{r,\mathrm{e}}^2)$ in counts\,arcsec$^{-2}$, where $t_\mathrm{exp}$ is the exposure time.

The treatment of the other S\'{e}rsic parameters, in particular the total emission and half-light radius, in the context of population studies is described in Section~\ref{sec:populations} below.

\subsection{Lensing substructure}

The ultimate goal of our method is to characterize the substructure population in strong lenses. Here we describe our procedure to model the substructure contribution to the lensing signal. Understanding the expected abundance of substructure in galaxies over a large range of epochs is a complex problem and an active ongoing area of research. Properties of individual subhalos (such as their density profiles) as well as those that describe their population (such as the mass and spatial distribution) are strongly affected by their host environment, and accurately modeling all aspects of subhalo evolution and environment is beyond the scope of this paper. Instead, we use a simplified description to model the substructure contribution in order to highlight the broad methodological points associated with the application of our method.

 $\Lambda$ Cold Dark Matter (\lcdm), often called the standard model of cosmology, predicts a scale-invariant power spectrum of primordial fluctuations and the existence of substructure over a broad range of masses with approximately equal contribution per logarithmic mass interval. We parameterize the distribution of subhalo masses $\mtwo$ in a given host halo of mass $\Mtwo$---the subhalo mass function---as a power law distribution with a linear dependence on the host halo mass,
\begin{equation}
\frac{\diff n}{\diff \log \frac {\mtwo}{m_{200,0}}} =
\begin{cases}
  \alpha \frac{\Mtwo}{M_{200,0}} \!\left(\!\frac{\mtwo}{m_{200,0}}\!\right)^{\!\beta} & \scriptstyle (m_\mathrm{200}^\mathrm{min} \leq \mtwo \leq m_\mathrm{200}^\mathrm{max}) \\
  0 & \scriptstyle (\text{else})\,,
\end{cases}
\label{eq:shmf}
\end{equation}
where $\alpha$ encodes the overall substructure abundance, with larger $\alpha$ corresponding to more substructure, and the slope $\beta < 0$ encodes the relative contribution of subhalos at different masses, with more negative $\beta$ corresponding to a steeper slope with more low-mass subhalos. $m_{200, 0}$ and $M_{200, 0}$ are arbitrary normalization factors.

Theory and simulations within the framework of \lcdm~predict a slope $\beta\approx-0.9$~\citep{0802.2265,0809.0898}, resulting in a nearly scale-invariant spectrum of subhalos, which we assume in our fiducial setup. We parameterize the overall subhalo abundance $\alpha$ through the mass fraction within the lensing galaxies contained in subhalos, $f_\mathrm{sub}$, defined as the fraction of the total dark matter halo mass contained in bound substructure in a given mass range:
\begin{equation}
f_\mathrm{sub} = \frac{\int_{m_\mathrm{200, min}}^{m_\mathrm{200, max}}\diff \mtwo\,\mtwo\,\frac{\diff n}{\diff \mtwo}}{M_\mathrm{200}} \,.
\end{equation}
For a given $\left\{f_\mathrm{sub},\beta\right\}$ and host halo mass $M_\mathrm{200}$, this can be used to determine $\alpha$ in Equation~\eqref{eq:shmf}. The linear scaling of the subhalo mass function with the host halo mass $\Mtwo$ in Equation~\eqref{eq:shmf} is additionally described in~\citet{2016MNRAS.457.1208H,2017MNRAS.469.1997D}. In our fiducial setups, we take the minimum and maximum subhalo masses to be $m_\mathrm{200, min} = 10^6\,\Msun$ and $m_\mathrm{200, max} = 0.01\,\,M_\mathrm{200}$~\citep{2017MNRAS.469.1997D,2018PhRvD..97l3002H} respectively, and corresponding fiducial substructure mass fraction in this range of 5\%, roughly consistent with observations in~\citet{2002ApJ...572...25D,2018PhRvD..97l3002H,2019arXiv190504182H}.

With all parameters of the subhalo mass function specified, the total number of subhalos $\mean{n}_{\mathrm{tot}}$ expected within the virial radius $\Rtwo$ of the host halo can be inferred as $\int_{m_\mathrm{200, min}}^{m_\mathrm{200, max}}\diff \mtwo\,\frac{\diff n}{\diff \mtwo}$. Strong lensing probes a region much smaller than this scale---the typical Einstein radii for the host deflector are much smaller than the virial radius of the host dark matter halos. In order to obtain the expected number of subhalos within the lensing observation's region of interest (ROI), we scale the total number of subhalos obtained from the above procedure by the ratio of projected mass within our region of interest $\theta_\textrm{ROI}$ and the host halo mass $\Mtwo$ as follows. We assume the subhalos to be distributed in number density following the host NFW dark matter profile. In this case, the enclosed mass function is $M_\mathrm{enc}(x) = \Mtwo\left[\ln(x/2) + \mathcal{F}(x)\right]$~\citep[\eg,][]{2001astro.ph..2341K}, where $x$ is the angular radius in units of the scale radius, $x\equiv \theta/\theta_\mathrm{s}$ and $\mathcal{F}(x)$ is given by Equation~\eqref{eq:Fx} above. The expected number of subhalos within our ROI is thus obtained as $\mean n_\mathrm{ROI} = \mean n_\mathrm{tot}\left[\ln(x_\mathrm{ROI}/2) + \mathcal{F}(x_\mathrm{ROI})\right]$. We conservatively take the lensing ROI to enclose a region of angular size twice the Einstein radius of the host halo, $\theta_\mathrm{ROI} = 2\cdot\theta_\mathrm{E}$.

Since strong lensing probes the line-of-sight distribution of subhalos within the host, their projected spatial distribution is approximately uniform within the lensing ROI~\citep{2017MNRAS.469.1997D}. We thus distribute subhalos uniformly within our ROI. The density profile of subhalos is assumed to be NFW and given by Equation~\eqref{eq:rhoNFW}, with associated lensing properties as described and the concentration inferred using the model in~\citet{2014MNRAS.442.2271S}.

We finally emphasize that we do not intent to capture all of the intricacies of the subhalo distribution, such as the effects of baryonic physics, tidal disruption of subhalos in proximity to the center of the host and redshift evolution of host as well as substructure properties. Although our description can be extended to take these effects into account (see Section~\ref{sec:extensions}), their precise characterization is still subject to large uncertainties, and our simple model above captures the essential physics for demonstration purposes.

\subsection{Observational considerations}
\label{sec:observations}

Our method is best-suited to analyzing a statistical sample of strong lenses, such as those that are expected to be obtained in the near future with optical telescopes like \Euclid and LSST, to quantify the effect of substructure. Given the challenges associated with the precise characterization of such a sample at the present time, we describe here the observational characteristics we assume in order to build up training and testing samples to validate our inference techniques.

We largely follow the description of~\citet{2015ApJ...811...20C} and use the associated \package{LensPop} package to characterize our mock observations. In particular, we use the nominal detector configuration for \Euclid, assuming a zero-point magnitude $m_\mathrm{AB} = 25.5$ in the single optical VIS passband, a $64\times64$ pixel grid with pixel size 0.1\,arcsec, a Gaussian point spread function (PSF) with FWHM 0.18\,arcsec, individual exposures with exposure time 1610\,s, and an isotropic sky background with magnitude 22.8\,arcsec$^{-2}$ in the detector passband.

These properties, in particular the exposure, sky background, and PSF shape, are expected to vary somewhat across the lens sample. Additionally, a given region may be imaged by multiple exposures over a range of color bands. Although such variations can be incorporated into our analysis, modeling these features is beyond the scope of this study. We comment on these extensions in Section~\ref{sec:extensions}.

\subsection{Population properties of the lens and source samples}
\label{sec:populations}

The fact that the strong lens population is expected to be dominated by higher-redshift ($z_\mathrm{s} \gtrsim1$) blue source galaxies lensed by intermediate-redshift ($z_\mathrm{l} \sim 0.5$--$1$) elliptical galaxies presents significant challenges for quantifying the lens population obtainable with future observations. Specifically, planned ground-based surveys like LSST and space telescopes like \Euclid present complementary challenges for delivering images of strong lensing systems suitable for substructure studies. LSST is expected to image in six bands, allowing for efficient separation between source and lens emission, but at the cost of lower resolution by virtue of being a ground-based instrument. \Euclid imaging is expected be higher in resolution but with a single optical passband (VIS). Near-IR imaging from WFIRST may deliver a high-resolution, multi-wavelength dataset that is more suitable for substructure studies, although potentially with different lens and source samples from those deliverable by optical telescopes.

In light of these uncertainties, we confine ourselves to a setting where the main methodological points can be made without detailed modeling of the detector capabilities and the deliverable lensing dataset, which is outside of the scope of the current paper. For concreteness, we simulate a sample of lenses with a simplified subset of host galaxy properties consistent with those deliverable by \Euclid as modeled by~\citet{2015ApJ...811...20C}. In particular, we assume spherical lenses, with ellipticity parameter $q=1$ in Equation~\eqref{eq:hostprofile}. We draw the central 1-D velocity dispersions $\sigma_v$ of host galaxies from a normal distribution with mean 225\,km\,s$^{-1}$ and standard deviation 50\,km\,s$^{-1}$. Following \citet{2018ApJ...859...96Z}, Equation~\eqref{eq:sigma_v_M_200_relation} is used to map the drawn $\sigma_v$ to a dark matter halo mass $\Mtwo$, and the host Einstein radius is analytically inferred with Equation~\eqref{eq:siethetae}.

\begin{figure*}
\centering
\includegraphics[width=1.\textwidth]{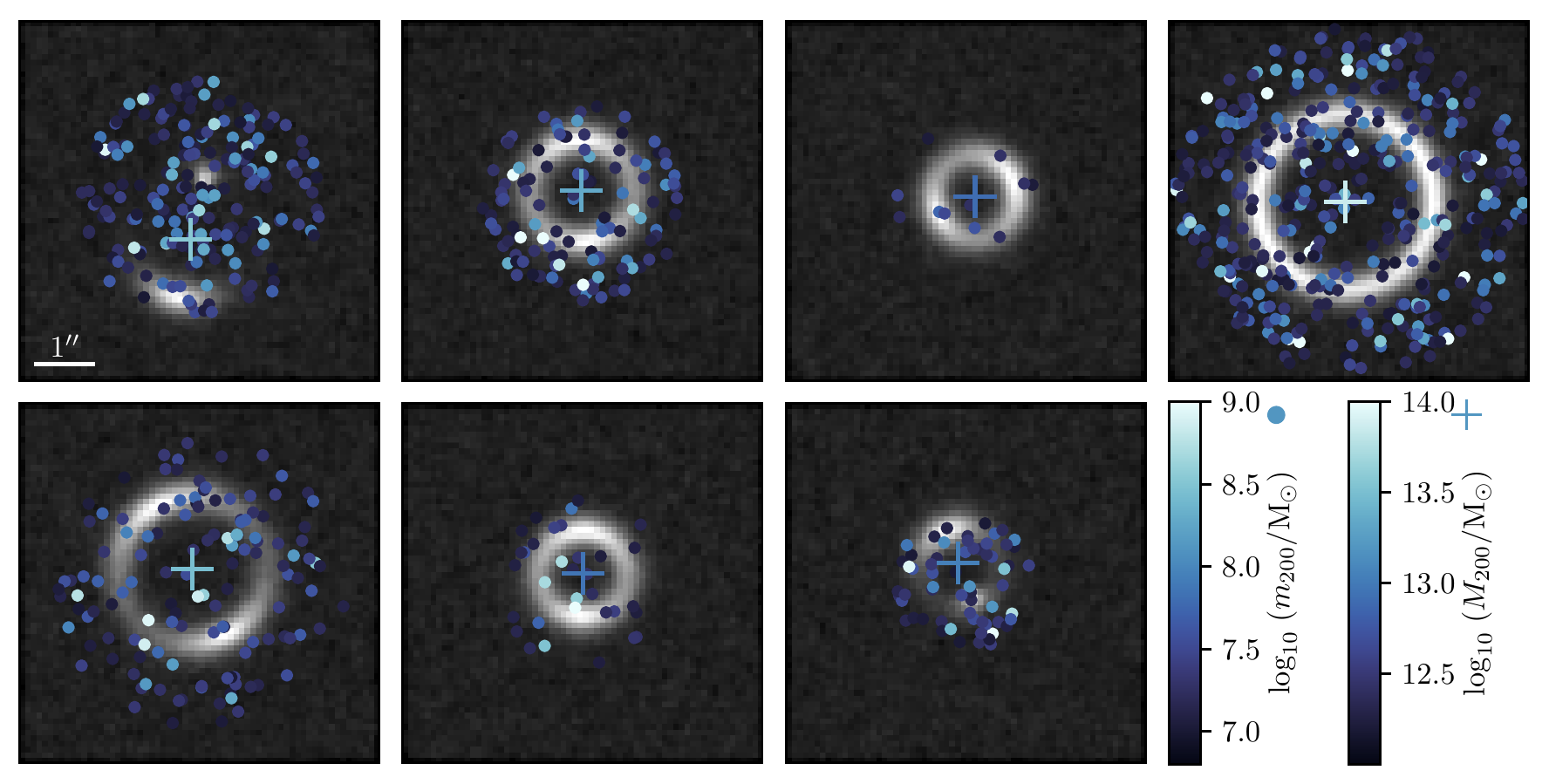}
\caption{A sample of simulated lenses. The foreground host galaxy is centered in each image and the cross markers show the (offset) position of the center of the source galaxy with the color corresponding to the virial mass of the host dark matter halo. The simulated subhalos are shown as dots, the color again indicates their masses. The greyscale images show the corresponding observed images. We show seven images randomly generated for $\fsub = 0.05$ and $\beta = -0.9$. \nblink{1_simulations}}
\label{fig:simulations}
\end{figure*}

We draw the lens redshifts $z_\mathrm{l}$ from a log-normal distribution with mean 0.56 and scatter 0.25 dex, discarding lenses with $z_\mathrm{l} > 1$ as these tend to have a small angular size over which substructure perturbations are relevant. The source redshift is fixed at $z_\mathrm{s} = 1.5$, its offsets $\Delta\theta_x$ and $\Delta\theta_y$ are drawn from a normal distribution with zero mean and standard deviation 0.2. These choices are consistent with the lens sample generated from the \package{LensPop} code packaged with~\citet{2015ApJ...811...20C}. We show a sample of simulated lensed images with these settings in Figure~\ref{fig:simulations}.

\section{Statistical formalism and simulation-based inference}
\label{sec:lfi-formalism}
Our goal is to infer the subhalo mass function parameters from a catalog of images of observed lenses. In this section we will describe the challenges of this inference problem and our approach of simulation-based inference. For simplicity, we will use a more abstract notation, distinguishing between three sets of quantities in the lensing system:
\begin{description}
  \item[Parameters of interest $\stattheta$] The vector $\stattheta = (\fsub, \beta)^T$ parameterizes the subhalo mass function given, and our goal is to infer their values.
  \item[Latent variables $z$] A vector of all other unobservable random variables in the simulator. These include the mass $M_{200}$, source-host offset $(\Delta\theta_x, \Delta\theta_y)$, and redshift $z_\mathrm{l}$ of the lens, the number of subhalos in the region of interest $n_\mathrm{ROI}$, the position $\mathbf{r}$ and mass $m_{200}$ of each subhalo, and the random variables related to the point spread function and Poisson fluctuations.
  \item[Observables $x$] The observed lens images.
\end{description}
Unfortunately, the same symbols are used with different meanings in astrophysics and statistics: note the difference between the parameters $\stattheta$ and the angular positions $\theta_x$, $\theta_y$ and the Einstein radius $\theta_\mathrm{E}$; between the latent variables $z$ and the redshifts $z_\mathrm{s}$, $z_\mathrm{l}$; and between the observed image $x$ and the argument of the NFW profile $M_\mathrm{enc}(x)$ and $\mathcal F(x)$ used in the last section.

As described above, we have implemented a simulator for the lensing process in the ``forward'' direction: for given parameters $\stattheta$, the simulator samples latent variables $z$ and finally observed images $x \sim p(x|\stattheta)$. Here $p(x|\stattheta)$ is the probability density or likelihood function of observing a lens image $x$ given parameters $\stattheta$. It can be schematically written as
\begin{equation}
 p(x|\stattheta) = \int\!\diff z \; p(x, z|\stattheta) \,,
 \label{eq:likelihood_latent}
\end{equation}
where we integrate over the latent variables $z$ and $p(x, z|\stattheta)$ is the joint likelihood of observables and latent variables:
\begin{multline}
  p(x,z | \stattheta)
  = p_\mathrm{host}(M_{200}, \Delta\theta_x, \Delta\theta_y, z_\mathrm{l}) \\
  \times  \pois(n_\mathrm{ROI} | \mean{n}_\mathrm{ROI}(\stattheta)) \prod_i^{n_\mathrm{ROI}} \Bigl[ p_m \! \left( m_{200, i} \middle| \stattheta \right) \; \uniform \left( \mathbf{r}_i \right) \Bigr] \\
  \times p_\mathrm{obs} ( x | f(M_{200}, \Delta\theta_x, \Delta\theta_y, z_\mathrm{l}; \{(m_{200, i},\mathbf{r}_i)\})) \,.
  \label{eq:joint_likelihood}
\end{multline}
Here $p_\mathrm{host}(M_{200}, \Delta\theta_x, \Delta\theta_y, z_\mathrm{l})$ is the distribution of the host halo parameters; $\mean{n}_\mathrm{ROI}(\stattheta)$ is the mean number of subhalos in the region of interest as a function of the parameters $\stattheta = (\fsub, \beta)^T$, while $n_\mathrm{ROI}$ is the actually realized number in the simulation; $m_{200, i}$ and $\mathbf{r}_i$ are the subhalo masses and positions; $p_m(m|\stattheta) = 1/n\, \diff n / \diff m_{200}$ is the normalized subhalo mass function given in Equation~\eqref{eq:shmf}; and in the last line $p_\mathrm{obs}$ is the probability of observing an image $x$ based on the true lensed image $f(z_\mathrm{l},\{(m_{200, i},r_i)\})$ taking into account Poisson fluctuations and detector response through the point spread function.

Standard frequentist and Bayesian inference methods rely on evaluating the likelihood function $p(x|\stattheta)$. Unfortunately, even in our somewhat simplified simulator each run of the simulation easily involves hundreds to thousands of latent variables, the integral in Equation~\eqref{eq:likelihood_latent} over this enormous space clearly cannot be computed explicitly. The likelihood function $p(x | \stattheta)$ is thus intractable, providing a major challenge for both frequentist and Bayesian inference. Similarly, inference with Markov Chain Monte Carlo (MCMC) methods based directly on the joint likelihood function $p(x,z | \stattheta)$ requires unfeasibly many samples before converging because the latent space is so large. Systems defined through a forward simulator that does not admit a tractable likelihood are known as ``implicit models'', inference techniques for this case as ``simulation-based inference'' or ``likelihood-free inference''.

One way to tackle this issue is to estimate the density for observables $x$ from samples from the simulator, where the latent variables $z$ are marginalized by the sampling procedure. But traditional density estimation techniques require reducing the dimensionality of $x$ with summary statistics $v(x)$, for instance based on power spectra~\citep{1403.2720,1506.01724,1707.04590,1710.03075,1809.00004,1806.07897,1808.03501}. The likelihood $p(v|\stattheta)$ in the space of summary statistics can either be explicitly estimated through density estimation techniques such as histograms, kernel density estimation, or Gaussian processes, or replaced by a rejection probability in an Approximate Bayesian Computation (ABC) technique~\citep{rubin1984}. Substructure inference in quasar and extended-arc lenses using ABC techniques was explored in~\citet{1712.04945} and~\citet{1702.00009}, respectively. While the compression to summary statistics makes the analysis tractable, it typically loses information and hence reduces the statistical power of the analysis.

Instead, the likelihood function or density can be approximated without any compression to summary statistics with a neural network, which has to be trained only once and can be evaluated efficiently for any parameter point and observed image. Similarly, one can train a neural network to estimate the likelihood \emph{ratio} for a fixed observation $x$ between two different hypotheses or parameter points. We will show how this turns the intractable integral in Equation~\eqref{eq:likelihood_latent} into a tractable minimization problem and amortizes the marginalization over $z$. This approach scales well to the expected large number of lenses expected in upcoming surveys~\citep{1001.2037,1003.5567,2015ApJ...811...20C}. Since the full image is used as input, there is no loss of information from a dimensionality reduction to summary statistics.

We use a simulation-based inference technique introduced in \citet{1805.00013,1805.00020,1805.12244} that extracts additional information from the simulation and uses it to improve the sample efficiency of the training of the neural network. Our inference strategy consists of four steps:
\begin{enumerate}
  \item During each run of the simulator, additional information that characterizes the subhalo population and lensing process is stored together with the simulated observed image.
  \item This information is used to train a neural network to approximate the likelihood ratio function.
  \item The neural network output is calibrated, ensuring that errors during training do not lead to incorrect inference results.
  \item The calibrated network output is then used in either a frequentist or Bayesian setting to perform inference.
\end{enumerate}
In the remainder of this section, we will explain these four steps in detail.

\subsection{Extracting additional information from the simulator}
\label{sec:lfi-gold}

In a first step, we generate training data by simulating a large number of observed lenses. For each lens, we first draw two parameter points from a proposal distribution, $\stattheta, \stattheta' \sim \pi(\stattheta)$. This proposal distribution should cover the region of interest in the parameter space, but does not have to be identical to the prior in a Bayesian inference setting, which allows us to be agnostic about the inference setup at this stage. Note that we use the term ``proposal distribution'' to avoid confusion with the prior, even though it is not a proposal distribution in the MCMC sense.

Next, the simulator is run for the parameter point $\stattheta$, generating an observed image $x \sim p(x|\stattheta)$. In addition, we calculate and save two quantities: the joint likelihood ratio
\begin{equation}
  r(x,z | \stattheta) = \frac {p(x,z | \stattheta)} {p_\mathrm{ref}(x,z)}
\end{equation}
and the joint score
\begin{equation}
  t(x, z | \stattheta) = \nabla_{\stattheta} \log p(x,z | \stattheta) \,.
\end{equation}
The joint likelihood ratio quantifies how much more or less likely a particular simulation chain including the latent variables $z$ is for the parameter point $\stattheta$ compared to a reference distribution
\begin{equation}
  \pref(x,z) = \int\!\diff\stattheta' \, \pi(\stattheta') \, p(x,z | \stattheta') \,,
\end{equation}
where we choose the marginal distribution of latent variables and observables corresponding to the proposal distribution $\pi(\stattheta)$. Unlike the distribution for a single reference parameter point, this marginal model has support for every potential outcome of the simulation~\citep{Hermans:2019ioj}. The joint score is the gradient of the joint log likelihood in model parameter space and quantifies if a particular simulation chain becomes more or less likely under infinitesimal changes of the parameters of interest. Both quantities depend on the latent variables of the simulation chain.

We compute the joint likelihood ratio and joint score with Equation~\eqref{eq:joint_likelihood}. Conveniently, the first and third line of that equation do not explicitly depend on the parameters of interest $\stattheta$ and cancel in the joint likelihood ratio and joint score; the remaining terms can be evaluated with little overhead to the simulation code. We also calculate the joint likelihood ratio $r(x,z|\stattheta')$ and the joint score $t(x,z|\stattheta')$ for the second parameter point $\stattheta'$ and store the parameter points $\stattheta$ and $\stattheta'$, the simulated image $x$, as well as the joint likelihood ratios and joint scores.

Our training samples consist of $10^6$ images, with parameter points chosen from a uniform range in $0.001 < \fsub < 0.2$ and $-1.5 < \beta < -0.5$.

\subsection{Machine learning}
\label{sec:lfi-ml}

How are the joint likelihood ratio and joint score, which are dependent on the latent variables $z$, useful for inference based on the likelihood function $p(x|\stattheta)$, which only depends on the observed lens images and the parameters of interest? Consider the functional
\begin{multline}
  L[g(x, \stattheta)] = \int\!\diff\stattheta\! \int\!\diff\stattheta'\! \int\!\diff x\! \int\!\diff z \; \pi(\stattheta) \; \pi(\stattheta') \; p(x,z|\stattheta) \\
    \times \Biggl[
    - s \log g  - (1 - s) \log (1 - g) - s' \log g'  - (1 - s') \log (1 - g') \\
    + \alpha \Bigl\{ \left| t - \nabla_\stattheta \log \tfrac{1 - g}g \Bigr|_{\stattheta}  \right|^2
    + \left| t' - \nabla_\stattheta \log \tfrac{1 - g}g \Bigr|_{\stattheta'} \right|^2 \Bigr\}
   \Biggr]  \,,
   \label{eq:alices_loss}
\end{multline}
where we abbreviate $s \equiv s(x,z|\stattheta) \equiv 1 / (1 + r(x,z|\stattheta))$,  $s' \equiv s(x,z|\stattheta') \equiv 1 / (1 + r(x,z|\stattheta'))$, $g \equiv g(x, \stattheta)$, \mbox{$g' \equiv g(x, \stattheta')$}, $t = t(x,z | \stattheta)$, and $t' \equiv t(x,z | \stattheta')$ for readability. Note that the test function $g(x, \stattheta)$ is a function of $x$ and $\stattheta$ only. The first two lines of Equation~\eqref{eq:alices_loss} are an improved version of the cross-entropy loss, in which the joint likelihood ratio is used to decrease the variance compared to the canonical cross-entropy~\citep{Stoye:2018ovl}. The last line adds gradient information, weighted by a hyperparameter $\alpha$.

As shown in \citet{Stoye:2018ovl}, this ``ALICES'' loss functional is minimized by the function
\begin{equation}
  g^*(x, \stattheta) \equiv \argmin_g L[g(x, \stattheta)] = \frac 1 {1 + r(x|\stattheta)} \,,
  \label{eq:loss_minimum}
\end{equation}
one-to-one with the likelihood ratio function
\begin{equation}
  r(x|\stattheta)
  \equiv \frac {p(x|\stattheta)} {p_\mathrm{ref}(x)}
  = \frac {1 - g^*(x, \stattheta)}{g^*(x, \stattheta)} \,.
\end{equation}
We demonstrate the minimization of this functional explicitly in Appendix~\ref{app:variation}. This means that if we can construct the functional in Equation~\eqref{eq:alices_loss} with the joint likelihood ratio and joint score extracted from the simulator and numerically minimize it, the resulting function lets us reconstruct the (otherwise intractable) likelihood ratio function $r(x|\stattheta)$! Essentially, this step lets us integrate out the dependence on latent variables $z$ from the joint likelihood ratio and score, but in a general, functional form that does not depend on a set of observed images.

This is why extraction of the joint likelihood ratio and joint score has been described with the analogy of ``mining gold'' from the simulator~\citep{1805.12244}---while calculating these quantities may require some effort and changes to the simulator code, through the minimization of a suitable functional they allow us to calculate the otherwise intractable likelihood ratio function.

In practice, we implement this minimization with machine learning. A neural network plays the role of the test function $g(x, \stattheta)$, the integrals in Equation~\eqref{eq:alices_loss} are approximated with a sum over training data sampled according to $\pi(\stattheta) \pi(\stattheta') p(x,z|\stattheta)$, and we minimize the loss numerically through a stochastic gradient descent algorithm. The neural network trained in this way provides an estimator $\hat{r}(x|\stattheta)$ of the likelihood ratio function that is exact in the limit of infinite training samples, sufficient network capacity, and efficient minimization. Note the ``parameterized'' structure of the network, in which a single neural network is trained to estimate the likelihood ratio over all of the parameter space, with the tested parameter point $\stattheta$ being an input to the network \citep{Cranmer:2015bka, Baldi:2016fzo}. This approach is more efficient than a point-by-point analysis of a grid of parameter points: it allows the network to ``borrow'' information from neighboring parameter points, benefit ting from the typically smooth structure of the parameter space.

Given the image nature of the lensing data, we choose a convolutional network architecture based on the ResNet-18 \citep{he2016deep} implementation in \package{PyTorch}~\citep{paszke2017automatic}. The parameters $\stattheta$ enter as additional inputs in the fully connected layers of the network. Compared to the original ResNet-18 architecture, we add another fully connected layer at the end to ensure that the relation between parameters of interest and image data can be modeled. All inputs are normalized to zero mean and unit variance. We train the networks by minimizing the loss in Equation~\eqref{eq:alices_loss} with $\alpha = 2 \cdot 10^{-3}$ over 100 epochs with a batch size of 128 using stochastic gradient descent with momentum~\citep{Qian:1999:MTG:307343.307376}, exponentially decaying the learning rate from 0.01 to 0.0001 with early stopping. We pretrain the model on data generated from a simplified version of the simulator, namely the ``fix'' scenario described in Appendix~\ref{app:simplified}. This architecture and hyperparameter configuration performed best during a rough hyperparameter scan, though for this proof-of-concept study we have not performed an exhaustive optimization.

\subsection{Calibration}
\label{sec:lfi-calibration}

With a finite data set and\,/\,or imperfect training, the neural network might not learn the likelihood ratio function $r(x|\stattheta)$ exactly, for instance due to limited training data or inefficient training. To make sure that our inference results are correct even in this case, we calibrate the network output with histograms~\citep{Cranmer:2015bka, 1805.00020}. For every parameter point $\stattheta$ that we want to test, we simulate a set of images $\{x\} \sim p(x|\stattheta)$ from this parameter point and calculate the network prediction $\hat{r} \equiv \hat{r}(x|\stattheta)$ for each image. We also simulate a set of images $\{x\} \sim p_{\mathrm{ref}}(x)$ from the reference model, again calculating the network prediction $\hat{r}$ for each lens. The calibrated likelihood ratio is then calculated from histograms of the network predictions as
\begin{equation}
  \hat{r}_\mathrm{cal}(x|\stattheta)
  = \frac {\hat{p}( \hat{r} | \stattheta )} {\hat{p}_\mathrm{ref}(\hat{r})}
\end{equation}
where the $\hat{p}(\cdot)$ denote probability densities estimated with univariate histograms.

This additional calibration stage comes with a certain computational cost that increases linearly with the number of evaluated parameter points. However, it guarantees that as long as the simulator accurately models the process, the inference results will be perfect or conservative, but not too optimistic, even if the neural network output is substantially different from the true likelihood ratio.

We will show results both without and with calibration. Where calibration is used, it is based on histograms with 50 bins, with bin boundaries determined automatically to match the distribution of likelihood ratios. Testing variations of the number of bins and bin boundary determination, we found that our results are robust to variations of the number of bins between 30 and 80.

\subsection{Inference}
\label{sec:lfi-inference}

After a neural network has been trained (and optionally calibrated) to estimate the likelihood ratio function, it provides the basic ingredient to both frequentist and Bayesian inference. Multiple observations can be combined in a straightforward way: since all lens images are assumed as identically distributed and independent (except for the common dependence on the population-level parameters), the combined likelihood of a set of images is given by the product of likelihood ratios for each individual lens,
\begin{equation}
  p_\mathrm{combined}(\{x\}|\theta) = \prod_{i} p(x_i | \theta) \,.
\end{equation}

For frequentist hypothesis tests, the most powerful test statistic to distinguish two parameter points $\theta_0$ and $\theta_1$ is the likelihood ratio~\citep{1933RSPTA.231..289N}
\begin{equation}
    \frac{p_\mathrm{combined}(\{x\}|\theta_0)}{p_\mathrm{combined}(\{x\}|\theta_1)} = \prod_{i} \frac{r(x_i | \theta_0)}{r(x_i | \theta_1)} \approx \prod_{i} \frac{\hat{r}(x_i | \theta_0)}{\hat{r}(x_i | \theta_1)} \,,
\end{equation}
where in the last step we have replaced the exact likelihood ratio with the estimation from the (calibrated) neural network. In addition, the asymptotic properties of the likelihood ratio allow us in many cases to directly translate a value of the likelihood ratio into a $p$-value and thus into exclusion limits at a given confidence level~\citep{Wilks:1938dza, Wald, Cowan:2010js}.

For Bayesian inference, note that we can write Bayes' theorem as
\begin{align}
  p(\stattheta | \{x_i\})
  &= \frac {p(\stattheta) \; \prod_i p(x_i | \stattheta)} {\int\!\diff \stattheta' \, p(\stattheta') \, \prod_i p(x_i | \stattheta')} \notag \\
  &= p(\stattheta) \Biggl[
    \int\!\diff\stattheta' \, p(\stattheta') \, \prod_i \frac {p(x_i | \stattheta')}{p(x_i | \stattheta)}
  \Biggr]^{-1} \notag \\
  &\approx p(\stattheta) \Biggl[
    \int\!\diff\stattheta' \, p(\stattheta') \, \prod_i \frac {\hat{r}(x_i | \stattheta')}{\hat{r}(x_i | \stattheta)}
  \Biggr]^{-1} \,,
  \label{eq:bayesian_post}
\end{align}
where $\{x_i\}$ is the set of observed lens images and $p(\stattheta)$ is the prior on the parameters of interest, which may be different from the proposal distribution $\pi(\stattheta)$ used during the generation of training data. The posterior can thus be directly calculated given an estimator $\hat{r}$, provided that the space of the parameters of interest is low-dimensional enough to calculate the integral, or with MCMC~\citep{Hermans:2019ioj} or variational inference techniques otherwise.

% \bigskip
While our approach to inference is strongly based on the ideas in \citet{1805.00013, 1805.00020, 1805.12244, Stoye:2018ovl}, there are some novel features in our analysis that we would like to highlight briefly. Unlike in those earlier papers, we use a marginal model based on the proposal distribution $\pi(\stattheta)$ as reference model in the denominator of the likelihood ratio, which substantially improves the numerical stability of the algorithm. This choice also allows us to include the ``flipped'' terms with $s'$ and $g'$ in the loss function in Equation~\eqref{eq:alices_loss}; we found that this new, improved version of the ALICES loss improves the sample efficiency of our algorithms. Both of these improvements are inspired by \citet{Hermans:2019ioj}. Finally, this is the first application of the ``gold mining'' idea to image data, the first combination with a convolutional network architecture, and the first use for Bayesian inference. Although machine learning-based methods have previously been proposed for inferring strong lensing host parameters~\citep{1708.08842,1708.08843,1808.00011} and for lensed source reconstruction~\citep{1901.01359}, this paper represents the first proposed application of machine learning for dark matter substructure inference in strong lenses and, as far as we are aware, for substructure inference in general.

\section{Results}
\label{sec:results}

\begin{figure*}
\centering
\includegraphics[width=1.\textwidth]{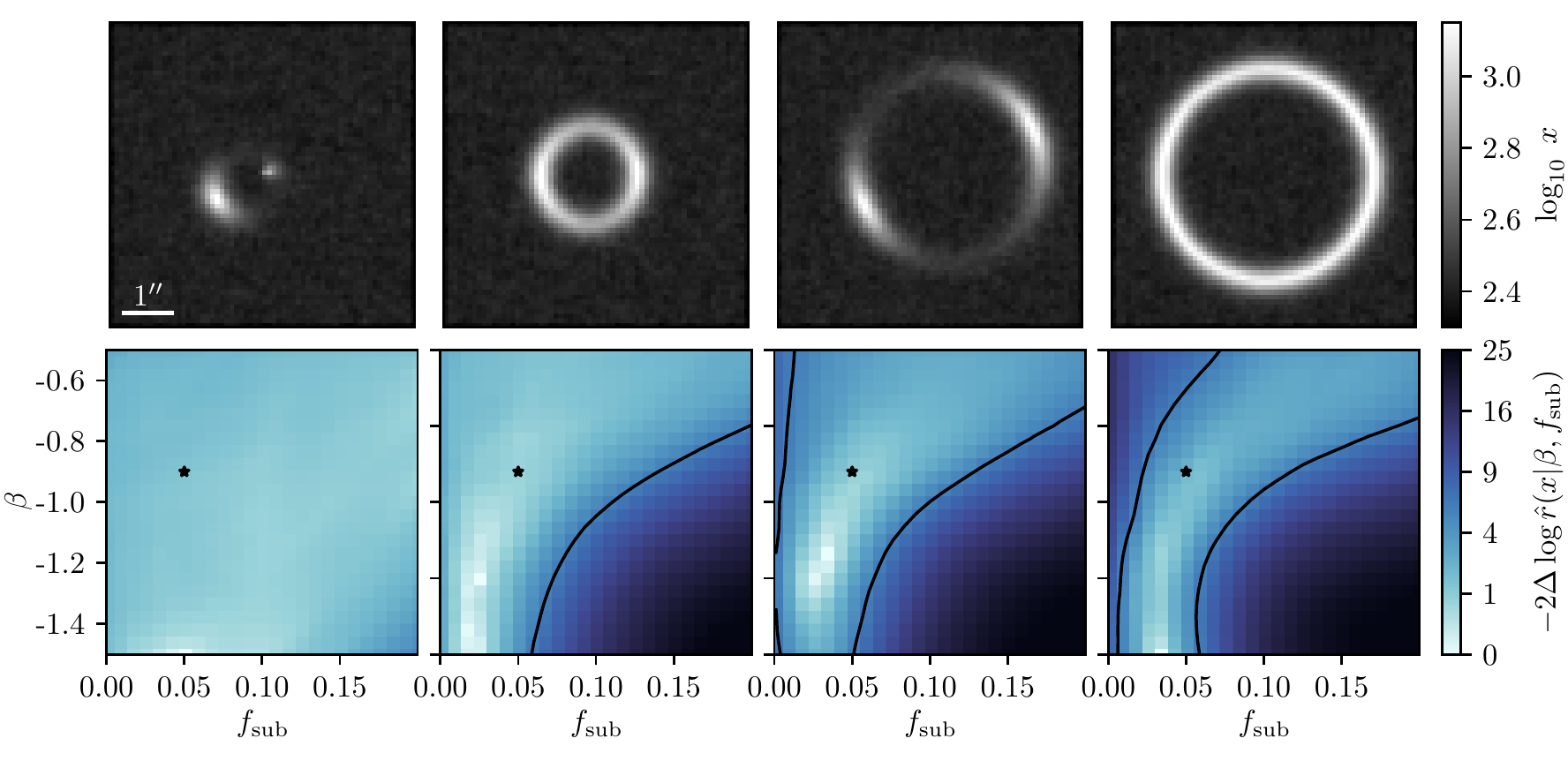}
\caption{Four simulated lens images (upper panels) and the corresponding likelihood ratio maps estimated by the network (lower panels, without calibration). The star marks the true point used to generate the images, the black line shows $95 \%$~CL contours in parameter space based on each image.  \nblink{2_inference_per_image}}
\label{fig:individual_predictions}
\end{figure*}

After training the neural network using the simulations described in Section~\ref{sec:lensing-formalism} and the formalism described in Section~\ref{sec:lfi-formalism}, we can run the inference step on a given set of images to extract the likelihood ratio estimates $\hat r(x | \stattheta)$ associated with the substructure parameters of interest $\{f_\mathrm{sub}, \beta\}$. We start by illustrating in Figure~\ref{fig:individual_predictions} inference on individual simulated lensed images realizing substructure corresponding to benchmark parameters $\beta = -0.9$ and $\fsub = 0.05$. The top row shows example simulated images, with the corresponding inferred 2-D likelihood surfaces shown in the bottom row. The true parameter point is marked with a star and the 95\% confidence level (CL) contours are shown.

Several interesting features can already be seen in these results. The 95\% CL contours contain the true parameter point, with the overall likelihood surface being strongly correlated with the corresponding image. A smaller projected surface area of the lensed arc, resulting from a smaller host halo or a larger offset between the host and source centers, generally results in a flatter likelihood surface. This is expected, since a smaller host galaxy will contain relatively less substructure, and a smaller host or larger relative offset will result in a smaller effective arc area over which the substructure can imprint itself. The first column of Figure~\ref{fig:individual_predictions} shows an example of such a system. In contrast, the last columns show a system with a relatively massive host and a small offset, producing a symmetric image with a larger effective arc surface area over which the effects of substructure can be discerned. This results in a ``peakier'' inferred likelihood surface, corresponding to a higher sensitivity to $\fsub$ and $\beta$. The second and third columns of Figure~\ref{fig:individual_predictions} correspond to systems with a small, centered and a large, offset halo respectively, and show intermediate sensitivity to substructure properties.

\begin{figure*}
  \centering
  \includegraphics[height=0.4\textwidth]{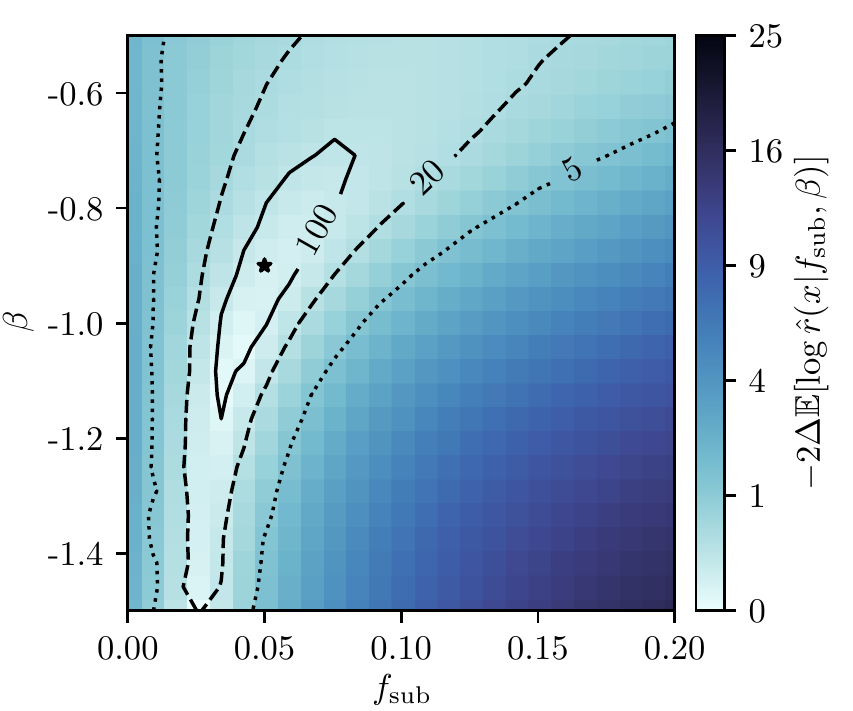}%
  \hspace*{0.075\textwidth}%
  \includegraphics[height=0.4\textwidth]{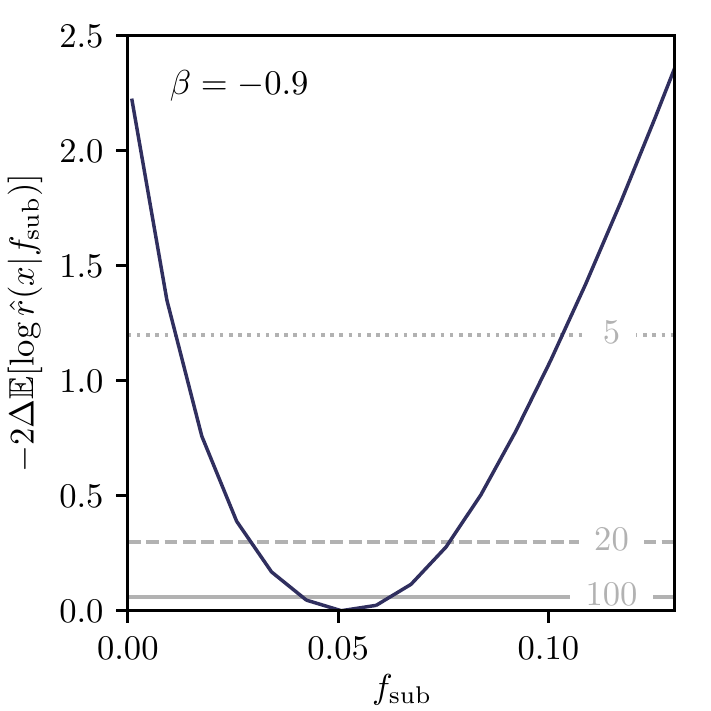}%
  \hspace*{0.052\textwidth}%
  \caption{The expected per-lens likelihood ratio map assuming $\beta = -0.9$ and $\fsub = 0.05$ in the two-dimensional parameter space (left) and along a one-dimensional slice at $\beta = -0.9$ (right). The lines show expected $95\%$~CL exclusion limits for 5 (dotted), 20 (dashed), and 100 (solid) observed lenses. While the colormap shows the network output without calibration, the lines include the calibration procedure described in Section~\ref{sec:lfi-calibration}. \nblink{3_6_expected_likelihood_map}}
  \label{fig:expected_likelihood}
\end{figure*}

In the spirit of stacking multiple observations, we next consider a simultaneous analysis of multiple lensed images. As discussed in Section~\ref{sec:lfi-inference}, the product of the likelihood maps of the individual images defines the appropriate test statistic. For the purpose of population-level inference, these two-dimensional likelihood maps are hence a good alternative way to define a probabilistic catalog over individual observations, avoiding the complications of prior dependence and of communicating a complicated trans-dimensional posterior. In the left panel of Figure~\ref{fig:expected_likelihood}, we show the expected log likelihood ratio surface per-image in the asymptotic limit, with the 1-D slice corresponding to $\beta = -0.9$ shown in the right panel. The 95\% CL expected exclusion limits for 5, 20, and 100 lenses are shown using the dotted, dashed, and solid lines respectively. The procedure can easily be extended to an arbitrarily large collection of lenses.

\begin{figure}
\centering
\includegraphics[height=0.4\textwidth]{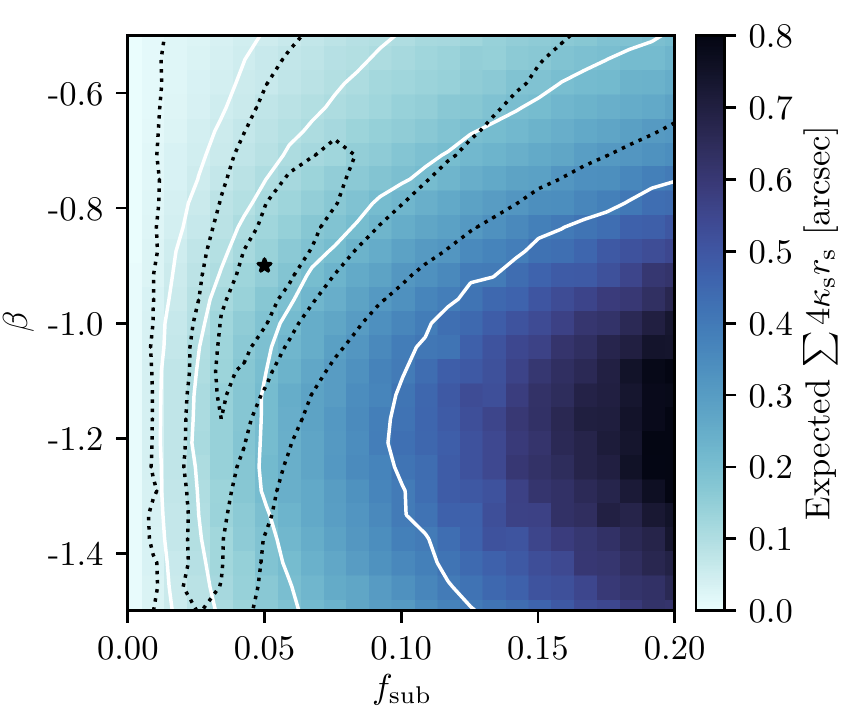}  % yeah, why is it?
\caption{Expected proxy for the total subhalo-induced deflection (see Equation~\eqref{eq:nfw_deflection}) as a function of $\fsub$ and $\beta$. The solid white lines show contours of constant deflection, while the dotted black lines show the expected exclusion limits from the left panel of Figure~\ref{fig:expected_likelihood}. \nblink{4_understand_the_banana}}
\label{fig:banana}
\end{figure}

We find that, at least within the simplifying assumptions of our simulator, an analysis of a few tens of lenses is already sensitive to the overall substructure abundance parameterized by $\fsub$. A larger observed lens sample provides a tighter constraint on substructure properties. Approximately 100 lens images are required to begin resolving $\beta$. The expected exclusion contours are centered around the true values, confirming that our inference methods yield an unbiased estimate of the underlying substructure properties. Note the ``banana'' shape of the expected exclusion limits, which approximately traces the total deflection contributed by substructure. We demonstrate this in Figure~\ref{fig:banana}, where we show a proxy for the total subhalo-induced deflection, $\sum_{\text{subhalos}} 4 \kappa_\mathrm{s} r_\mathrm{s}$, equal to the space-independent part of Equation~\eqref{eq:nfw_deflection}, and compare it to the expected exclusion limits. In our particular substructure scenario, this proxy can be shown to approximately scale like $\sum_{\text{subhalos}} m_{200}^{2/3}$. We note that this comparison is schematic, as the subtle effects of substructure over a wide range of masses cannot be quantified through a single number (here, the total deflection).

\begin{figure*}
\centering
\includegraphics[height=0.4\textwidth]{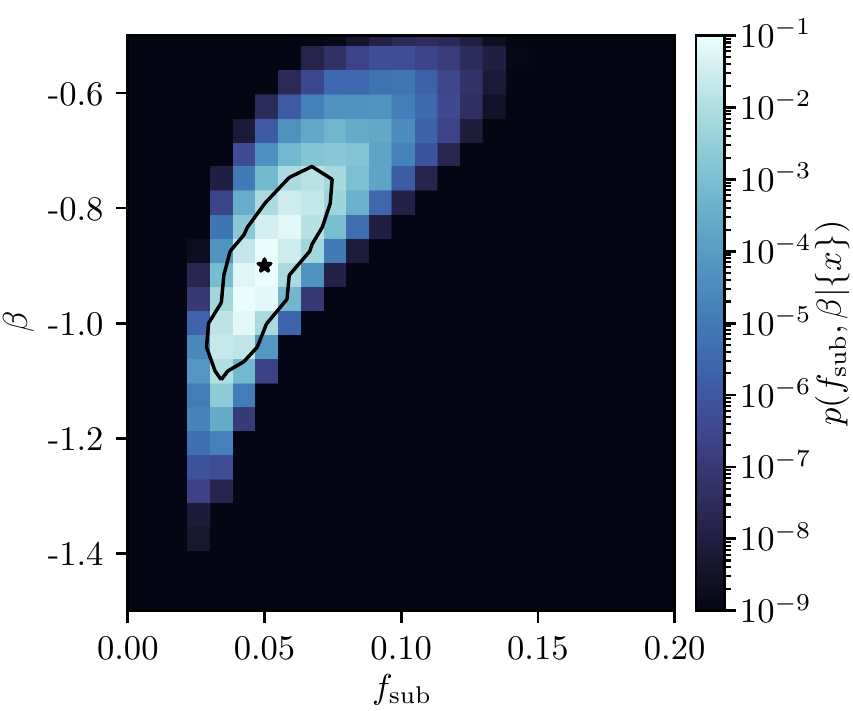}%
\hspace*{0.075\textwidth}%
\includegraphics[height=0.4\textwidth]{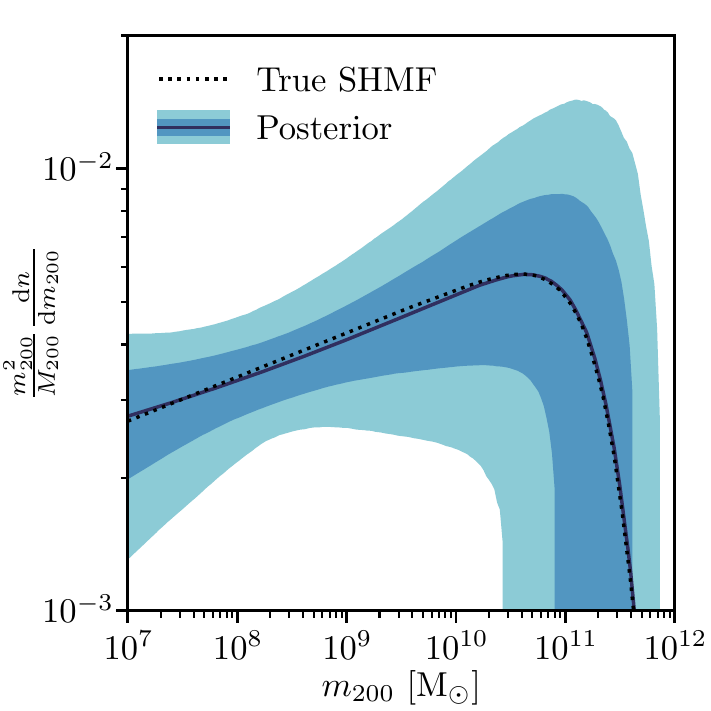}%
\hspace*{0.052\textwidth}%
\caption{Left: Expected posterior and $95\%$ credible region on subhalo mass function parameters for 100 observed lenses. The mock observations are generated for $\fsub = 0.05$ and $\beta = -0.9$. We assume a uniform prior on $\fsub$ and a Gaussian prior with mean $-0.9$ and standard deviation $0.1$ for $\beta$. \nblink{5_Bayesian_inference} Right: Corresponding inferred subhalo mass function, marginalized over the host halo properties. We show the point-wise mean (solid line) as well as point-wise 68\,/\,95\% credible intervals (cyan\,/\,blue bands) of the subhalo mass function per host halo mass. The true simulated subhalo mass function for $\fsub = 0.05$ and $\beta = -0.9$ is shown as the dotted line. An animation showing the evolution of the substructure parameter and mass function posteriors as the number of lenses analyzed is increased can be found at \animlink{live_inference_with_images_reverse_small}. \nblink{5_posterior_SHMF}}
\label{fig:bayesian_post}
\end{figure*}

With the likelihood ratio in hand, Equation~\eqref{eq:bayesian_post} easily admits a Bayesian interpretation. In the left panel of Figure~\ref{fig:bayesian_post} we show the posterior for 100 lenses derived from the expected likelihood ratio results, assuming a Gaussian prior with mean $-0.9$ and standard deviation $0.1$ on the slope $\beta$. This choice is intended to capture a prior expectation on the subhalo mass function slope consistent with the Cold Dark Matter scenario~\citep[\eg,][]{0802.2265,0809.0898}. As expected from the likelihood maps, we find a posterior density peaked around the true point.

The corresponding inferred subhalo mass function (SHMF) per host halo mass, marginalized over the host halo properties, is shown in the right panel of Figure~\ref{fig:bayesian_post}. We show the point-wise mean (solid line) and 68\,/\,95\% credible intervals (cyan and blue bands), where the point-wise quantities are defined as the mean and respective quantiles of the subhalo mass function posterior evaluated at a given mass point. A comparison with the true simulated subhalo mass function (dotted line, also marginalized over the host halo properties) shows excellent agreement.

\section{Extensions}
\label{sec:extensions}
For the proof-of-concept analysis presented here our lensing simulation makes a number of simplifying assumptions in order to highlight the broad methodological points in a computationally tractable setting. An application of our method to real lensing data will invariably require modifications to our simulation and inference pipelines to account for the vast physical diversity in host and source galaxy morphologies, as well as ways to deal with more realistic detector response. Modeling substructure in a more involved setting than presented here (\eg, to account for tidal evolution and/or suppression of small-scale structure), and accounting for substructure along the line of sight is also desired. We will now discuss these features and comment on how they might affect our pipeline and the results presented here, leaving implementation and application to real lensing data to future work.

First, we currently fix all properties of the background source as described in Section~\ref{sec:source}. It is straightforward to instead draw and marginalize over the parameters associated with a chosen parameterization for the source light distribution, with Gaussian and S\'{e}rsic~\citep{1963BAAA....6...41S} profile models being common choices. For high-fidelity images (\eg, those obtainable by targeted followups or interferometric imaging) more complicated features in the background galaxies such as outflows may not be adequately captured by such a parameterization and could introduce degeneracies with the effects of substructure. Alternative parameterizations using shapelet basis sets~\citep{1504.07629,1505.00198,1803.09746}, and methods based on regularized linear inversion on grids~\citep{2003ApJ...590..673W,astro-ph/0601493,1408.6297,1708.07377} have been introduced as ways to model more complicated source features. For our purposes, generative\,/\,data-driven modeling of background galaxies could easily be interfaced with our pipeline to account for the variation in structure of the background sources~\citep{1901.01359}.

Similarly, the host lens (and associated host dark matter halo) model can be made more realistic by relaxing the restriction to spherical host halos and including more complicated profiles than the Singular Isothermal Sphere considered here, drawing and marginalizing over additional host parameters as required. External shear, which models the fact that the local large-scale structure environment of the host galaxy can induce an additional overall deflection field in a preferred direction, can similarly be parameterized~\citep[\eg,][]{astro-ph/9610163,1997MNRAS.292..673S} and marginalized over.

A realistic simulator should also model the dynamical evolution of subhalos~\citep{2017MNRAS.469.1997D}. Effects associated with tidal disruption due to the large gradient of the galactic potential towards the center of the host galaxy are expected to deplete the fraction of mass bound in substructures there, leading to a depressed overall subhalo abundance~\citep{2016MNRAS.457.1208H} with profile properties (\eg, concentration~\citep{1603.04057} and a truncation radius~\citep{0705.0682}) that depend on the distance from the host center. This could easily be implemented within our framework by drawing 3-D positions for the subhalos from the host center and assigning properties consistent with more involved modeling. Our subhalo mass function in Equation~\eqref{eq:shmf} is independent of the lens redshift, but can easily be extended to include this dependency~\citep{2017MNRAS.469.1997D,2018PhRvD..97l3002H}. A more complicated dependence on the host halo than the linear one assumed in Equation~\eqref{eq:shmf} is also easily admitted.

All of these effects are straightforward to implement in our setup and only require modifications to the simulation code. The inference algorithm is unaffected; since these extensions do not explicitly depend on the parameters of interest, the likelihood terms associated with them cancel in the calculation of the joint likelihood ratio and the joint score. Nevertheless, these changes affect the final observed image and therefore also the true likelihood function; the variance of the joint likelihood ratio and score could therefore increase, requiring larger training samples before the network converges to the correct likelihood ratio function.

With these extensions, the redshift of the background source and the lens will play a more important role. Since these redshifts can potentially be measured through spectroscopic follow-up observations, it is likely that we can improve the performance of the inference algorithm by using this information. We can treat both the source and lens redshift, potentially with added uncertainty to model measurement noise, as additional observables. The input to the neural networks then consists of the observed lens image, the measured (potentially noisy) redshifts, and the tested parameter point. Except for a simple modification of the network architecture, the inference algorithm remains unchanged.

Including line-of-sight substructure can be somewhat more involved, since it necessitates the introduction of a separate line-of-sight halo mass function~\citep{1610.01599,1710.05029,1901.11031,2019arXiv190504182H}. Depending on the specific model (and whether foreground substructure is treated as a nuisance effect or additional signal to be leveraged) its parameters could depend on the parameters of interest, which would require a modification of the calculation of the joint likelihood ratio and joint score. Structurally this is identical to our current modeling of subhalos within the lens. Since the abundance of foreground substructure is expected to be at most comparable to the substructure within the lensing galaxy (depending on the source redshift), we expect that these additional factors in the joint likelihood ratio and joint score will not slow down the overall simulation significantly, and will not increase the variance of the inference techniques too much while having the potential to improve the overall sensitivity of the analysis to substructure abundance in the Universe.

Modification to the subhalo mass function parameterization that we have considered may be desirable for constraining specific particle physics scenarios. For example, warm dark matter introduces a lower cutoff scale in the subhalo mass function~\citep{2001ApJ...556...93B} which can be parameterized and mapped onto the dark matter mass~\citep{1112.0330,1308.1399,1512.06507,1702.00009}. This would also require a straightforward modification of the joint likelihood ratio and joint score calculation depending on the specific parameterization.

It is expected that a sample of strong lenses will include image-to-image variations on the exposure, sky background, and detector effects like the point spread function depending on the specific scanning strategy of the observatory. The sky background can be marginalized over as usual. Rather than treating the exposure and PSF model as nuisance parameters, passing them as additional a priori known inputs to the network in addition to normalizing the network input to unit exposure is likely to improve performance. Multiple color bands can easily be modeled and included as inputs to the neural network as different color channels, something that is commonly done when using the ResNet architecture we consider. This can substantially improve discrimination between light from the source, host, and sky background which tend to have a degree of separation in color space.

In addition to decreasing the sample efficiency, we expect that inclusion of the features discussed here---more complicated host and source profiles, as well as inclusion of line-of-sight substructure and external shear---will also degrade the overall sensitivity of our method to the properties of substructure. We explicitly demonstrate this in Appendix~\ref{app:simplified}, where we show that tighter constraints on substructure properties are obtained for scenarios where fewer degrees of freedom are associated with the host lens. This arises from the fact that the additional degrees of freedom can introduce features in the lensing image that are degenerate with the effects of substructure. A realistic and conservative substructure analysis must therefore necessarily take these extensions into account. While including these in our simulation and inference code is feasible, the detailed modeling is beyond the scope of the current paper. We thus leave the implementation of these features and application to real lensing data to future work.

\section{Conclusions}
\label{sec:conclusions}
Strong lensing offers a unique way to probe the properties and distribution of dark matter on sub-galactic scales through the subtle imprint of substructure on lensed arcs. The high dimensionality of the underlying latent space characterizing substructure poses a significant challenge, however. In this paper, we have introduced a novel simulation-based inference technique based on the ideas introduced in~\citet{Cranmer:2015bka, 1805.00013, 1805.00020, 1805.12244, Stoye:2018ovl} for inferring high-level population properties characterizing the distribution of substructure in an ensemble of galaxy-galaxy strong lenses and demonstrated its feasibility through proof-of-principle examples.

Our results on simulated data demonstrate that this method, based on calibrated likelihood ratio estimators with a machine learning back end, offers a promising way to analyze extended-arc strong lensing images with the goal of inferring properties of dark matter substructure. Our proposed method offers several combined advantages over established techniques. In probing the collective effect of a large number of low-mass, sub-threshold subhalos it can offer sensitivity to the faint end of the subhalo mass function where deviations from the concordance \lcdm paradigm and the effects of new physics are most likely to be expressed. It can naturally be applied to perform fast, principled, and concurrent analyses of a large sample of strong lenses that share a common set of hyperparameters describing the underlying substructure population properties. By efficiently marginalizing out the individual subhalo properties and directly inferring the population-level parameters of interest we are able to sidestep the more expensive two-step procedure of characterizing individual subhalos before using them to infer higher-level population parameters. Population-level likelihood scans for individual images are thus a suitable alternative to probabilistic catalogs over subhalos, avoiding both prior dependence as well as the logistical complexity of communicating a complicated trans-dimensional posterior. Furthermore, rigorous selection of lensing images out of a large sample is not necessary within our framework since images with a smaller effective arc area or low overall fidelity simply do not contribute significantly to the simultaneous substructure analysis, and non-detections are just as valuable as detections. Finally, our analysis is performed at the level of image data without incurring loss of information associated with dimensionality reduction.

Although we have focused on a simple proof-of-principle example in this paper, extensions to more realistic scenarios---including more complicated descriptions of the host, source, and substructure populations---are easily admitted within our framework.  The flexibility of the proposed method allows for applications beyond substructure population inference as well. For example, a large lens sample can be used to perform cosmological parameter estimation while accounting for substructure effects and in particular to independently constrain the Hubble constant~\citep{1907.02533,1907.04869} through its dependence on the angular diameter distance scales in lensing systems. Given the observed tension between early- and late-time probes of the Hubble constant~\citep{1807.06209,1903.07603,1907.04869}, the possibility of a subtle degeneracy between the effects of substructure and those due to variation of cosmological parameters further motivates the study and inclusion of substructure effects in measurements relying on the analysis of strong gravitational lenses. In the spirit of \citet{Alsing:2017var}, our methods can also be  used to learn powerful summary statistics~\citep{1805.12244}.

We are currently at the dawn of a new era in observational cosmology, when ongoing and upcoming surveys---\eg, DES, LSST, \Euclid, and WFIRST---are expected to discover and deliver images of thousands of strong lensing systems. These will harbor the subtle imprint of dark matter substructure, whose characterization could hold the key to unveiling the particle nature of dark matter. In this paper, we have introduced a powerful machine learning-based method that can be used to uncover the properties of small-scale structure within these lenses and in the Universe at large. The techniques presented have the potential to maximize the information that can be extracted from a complex lens sample and zero in on signatures of new physics.

The code used to obtain the results in this paper is available at \url{https://github.com/smsharma/mining-for-substructure-lens}\href{https://github.com/smsharma/mining-for-substructure-lens}~\githubmaster.

\vspace{-5mm}

\acknowledgments
We thank Simon Birrer, Christopher Fassnacht, Daniel Gilman, Siavash Golkar, and Neal Weiner for useful conversations. SM thanks Laura Chang for collaboration at the early stages of this work. JB and KC are partially supported by NSF awards ACI-1450310, OAC-1836650, and OAC-1841471, and the Moore-Sloan Data Science Environment at NYU. KC is also supported through the NSF grant PHY-1505463. JH thanks the F.R.S.-FNRS for his FRIA scholarship. SM is supported by the NSF CAREER grant PHY-1554858, NSF grants PHY-1620727 and PHY-1915409, and the Simons Foundation. This work was also supported through the NYU IT High Performance Computing resources, services, and staff expertise. This research has made use of NASA's Astrophysics Data System.
\software{
\package{Astropy} \citep{2013A&A...558A..33A,2018AJ....156..123A},
\package{IPython} \citep{PER-GRA:2007},
\package{Jupyter} \citep{Kluyver2016JupyterN},
\package{LensPop} \citep{2015ApJ...811...20C},
\package{MadMiner} \citep{Brehmer:2019xox},
\package{Matplotlib} \citep{Hunter:2007},
\package{NumPy} \citep{numpy:2011},
\package{Palettable} \citep{palettable},
\package{PyTorch} \citep{paszke2017automatic},
\package{SciPy} \citep{Jones:2001ab}.
}

\newpage

\appendix

\section{Minimum of the loss functional}
\label{app:variation}

A central step in our inference technique is numerically minimizing the functional $L[g(x, \stattheta)]$ given in Equation~\eqref{eq:alices_loss} to obtain an estimator for the likelihood ratio function. Here we will use calculus of variation to explicitly show that the solution given in Equation~\eqref{eq:loss_minimum} in fact minimizes this loss, closely following~\citet{1805.00020, Stoye:2018ovl}.

First consider the case of $\alpha = 0$, \ie the functional
\begin{align}
  L[g(x, \stattheta)]
  &= \int\!\!\diff\stattheta\!\!\int\!\!\diff\stattheta'\!\!\int\!\!\diff x\!\! \int\!\!\diff z \, \pi(\stattheta) \pi(\stattheta') p(x,z|\stattheta)
  \Bigl(- s \log g  - (1 - s) \log (1 - g) - s' \log g'  - (1 - s') \log (1 - g') \Bigr) \notag \\
  &= \int\!\!\diff\stattheta\!\!\int\!\!\diff x
  \underbrace{ \Biggl[
    \int\!\!\diff z \, \pi(\stattheta) \, \Bigl( p(x,z|\stattheta) + p_{\mathrm{ref}}(x,z) \Bigr)
    \Bigl(- s \log g  - (1 - s) \log (1 - g) \Bigr)
  \Biggr] }_{
  \equiv F(x,\stattheta)
  } \,,
\end{align}
where we use the shorthand notation $s \equiv s(x,z|\stattheta) \equiv 1 / (1 + r(x,z|\stattheta))$,  $s' \equiv s(x,z|\stattheta') \equiv 1 / (1 + r(x,z|\stattheta'))$, $g \equiv g(x, \stattheta)$, $g' \equiv g(x, \stattheta')$. The function $g^*(x|\stattheta)$ that minimizes  this functional has to satisfy
\begin{equation}
  0 \stackrel{!}{=} \frac {\delta F}{\delta g} \Biggr|_{g^*}
  =  \int\!\!\diff z \, \pi(\stattheta) \Bigl( p(x,z|\stattheta) + p_{\mathrm{ref}}(x,z) \Bigr) \Bigl( - \frac s {g^*} + \frac {1-s}{1-g^*} \Bigr)
\end{equation}
As long as $\pi(\stattheta) > 0$, this is equivalent to
\begin{equation}
  (1-g^*) \int\!\diff z \Bigl( p(x,z|\stattheta) + p_{\mathrm{ref}}(x,z) \Bigr) s
  = g^* \int\!\diff z \Bigl( p(x,z|\stattheta) + p_{\mathrm{ref}}(x,z) \Bigr) (1-s)
\end{equation}
and finally
\begin{align}
  g^*(x | \stattheta)
  &= \frac {\int\!\diff z \, \Bigl(p(x,z|\stattheta)+\pref(x,z)\Bigr) s(x,z|\stattheta)} {\int \! \diff z \, \Bigl(p(x,z|\stattheta)+\pref(x,z)\Bigr)} \notag \\
  &= \frac
  {\int\!\diff z \, \Bigl(p(x,z|\stattheta) + \pref(x,z)\Bigr) \frac{1}{1+p(x,z|\stattheta)/p_{\mathrm{ref}}(x,z)}}
  {\int\!\diff z \, \Bigl(p(x,z|\stattheta) + \pref(x,z)\Bigr)} \notag \\
  &= \frac {\pref(x)} {p(x|\stattheta) + \pref(x)}
  = \frac{1}
  {1 + r(x|\stattheta)} \,,
\end{align}
in agreement with Equation~\eqref{eq:loss_minimum}. Note that this result is independent of the choice of $\pi(\stattheta)$, as long as this proposal distribution has support at all relevant parameter points.

Similarly it can be shown that the gradient term in the loss functional weighted by $\alpha$ is minimized when the gradient of the log likelihood ratio estimated by the neural network is equal to the true score,
\begin{equation}
  \nabla_\stattheta \log \hat{r}(x|\stattheta) \equiv \nabla_\stattheta \log \frac {1 - g^*(x, \stattheta)}{g^*(x, \stattheta)} = \nabla_\stattheta \log r(x|\stattheta) \,.
\end{equation}
We refer the reader to \citet{1805.00020} for the derivation. While not strictly necessary for the inference technique, including this term in the loss function substantially improves the sample efficiency of the algorithm, similar to how gradient information makes any fit converge faster.

\section{Simplified scenarios}
\label{app:simplified}

\begin{figure*}
\centering
\includegraphics[height=0.4\textwidth]{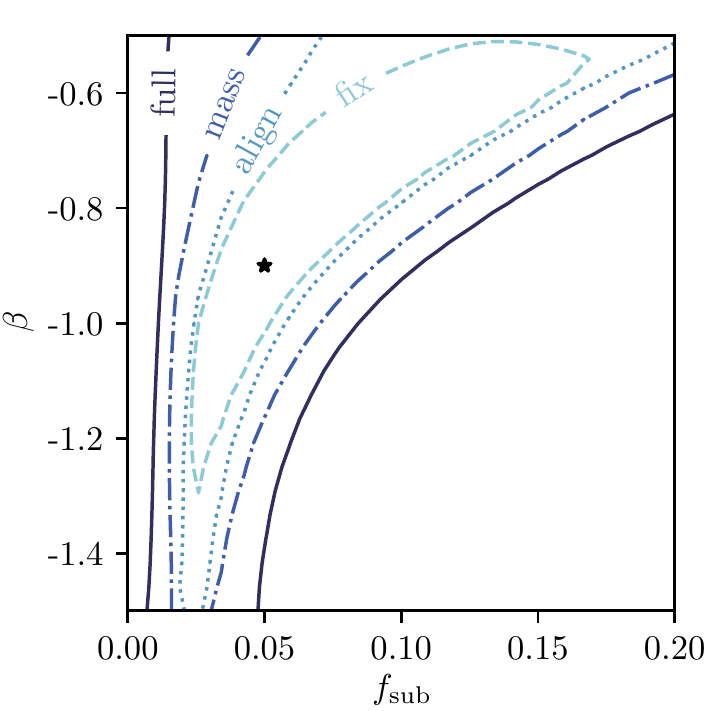}%%
\hspace*{0.052\textwidth}\hspace*{0.075\textwidth}%
\includegraphics[height=0.4\textwidth]{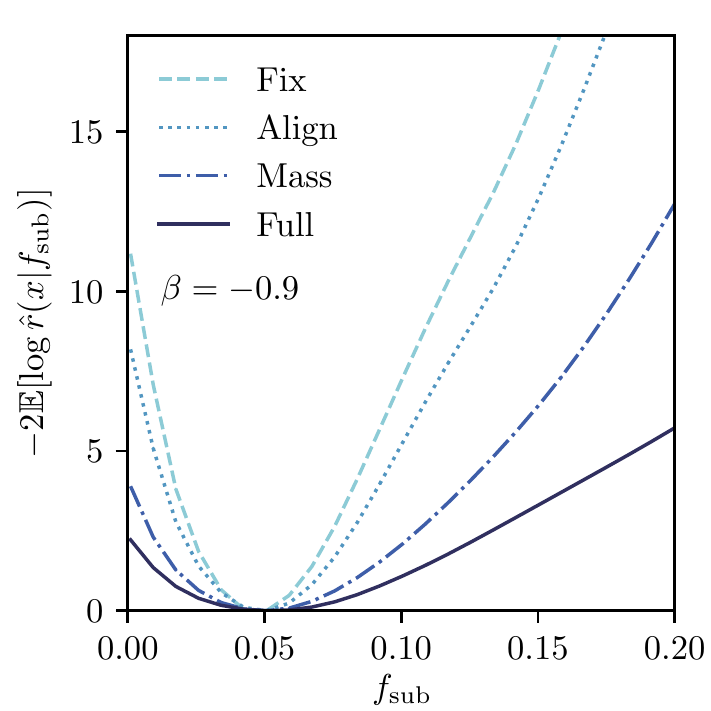}%
\hspace*{0.052\textwidth}%
\caption{Left: The expected $95\%$~CL exclusion limits for 5 observed lenses for four different levels of complexity of the simulator. Right: The expected likelihood ratio along a one-dimensional slice through the parameter space at $\beta = -0.9$ for the same four simulator scenarios. In both panels we compare the ``full'' simulator discussed in Section~\ref{sec:lensing-formalism}, a scenario in which the host mass is varied but the offset relative to the source is fixed at zero (``mass''), a case in which the source offset is varied but the host halo mass is fixed (``align''), and a toy scenario in which both the offset and the mass of the host halo are fixed (``fix''). The data was generated for $\beta = -0.9$ and $\fsub = 0.05$. \nblink{3_6_expected_likelihood_map}}
\label{fig:scenarios}
\end{figure*}

In order to validate our setup and to disentangle the impact of different latent variables on the inference results we consider three additional versions of our simulation. In the simplest one, which we call ``fix'', all source and host properties are fixed to particular value, including the host halo mass and the offset between source and lens, which is set to zero. In the ``align'' scenario we relax the restriction on the source offset variables $\Delta\theta_x$ and $\Delta\theta_y$ and draw them from a Gaussian as described in Section~\ref{sec:lensing-formalism}. In the ``mass'' version, on the other hand, the offset is fixed at zero, but the host halo mass is drawn from a distribution as described above. We train separate neural networks on lens images generated in these three scenarios and calculated likelihood maps as described in Section~\ref{sec:lfi-formalism}, although to save computation time we do not perform a calibration procedure.

The expected confidence limits for 5 observed lens images in the three simplified scenarios and our ``full'' setup are compared in Figure~\ref{fig:scenarios}. As expected, the more latent variables we keep fixed, the more the inference technique becomes more sensitive to the parameters of interest. In particular fixing the source-host alignment substantially increases the strength of the expected limits.

\bibliographystyle{aasjournal-mod}
\bibliography{lensing-lfi}

\end{document}

%% file: definitions.tex
\newcommand{\animicon}{{\color{xlinkcolor}\faPlayCircle}\xspace}
\newcommand{\nbicon}{{\color{xlinkcolor}\faFileCodeO}\xspace}

\newcommand{\animlink}[1]{\href{https://github.com/smsharma/mining-for-substructure-lens/blob/arXiv-v2/figures/#1.gif}{\animicon}}
\newcommand{\nblink}[1]{\href{https://github.com/smsharma/mining-for-substructure-lens/blob/arXiv-v2/notebooks/#1.ipynb}{\nbicon}}
\newcommand{\githubmaster}{\href{https://github.com/smsharma/mining-for-substructure-lens}{\faGithub}\xspace}

\newcommand{\package}[1]{\textsl{#1}\xspace}
\newcommand{\Euclid}{\textsl{Euclid}\xspace}
\newcommand{\lcdm}{$\Lambda$CDM\xspace}

\newcommand{\Msun}{\textrm{M}_\odot}
\newcommand{\kmps}{\textrm{km\,s}^{-1}}

\newcommand{\mtwo}{m_{200}}

\newcommand{\Mtwo}{M_{200}}

\newcommand{\Rtwo}{R_{200}}

\newcommand{\eg}{{e.\,g.}\xspace}
\newcommand{\ie}{{i.\,e.}\xspace}
\newcommand{\diff}{\mathrm{d}}

\newcommand{\fsub}{f_{\mathrm{sub}}}

\newcommand{\pref}{p_{\mathrm{ref}}}
\newcommand{\stattheta}{\vartheta}
\newcommand{\mean}[1]{\makebox[0pt]{$\phantom{#1}\overline{\phantom{#1}}$}#1}

\DeclareMathOperator*{\argmin}{arg\,min}
\DeclareMathOperator{\pois}{Pois}
\DeclareMathOperator{\uniform}{Uniform}